# Release Angle for Attaining Maximum Distance in the Soccer Throw-in

NICHOLAS P. LINTHORNE and DAVID J. EVERETT

School of Sport and Education, Brunel University, United Kingdom

## ABSTRACT

We investigated the release angle that maximises the distance attained in a long soccer throw-in. One male soccer player performed maximum-effort throws using release angles of between 10 and 60º, and the throws were analysed using 2-D video. The player's optimum release angle was calculated by substituting mathematical expressions for the measured relations between release speed, release height, and release angle into the equations for the flight of a spherical projectile. We found that the musculoskeletal structure of the player's body had a strong influence on the optimum release angle. When using low release angles the player released the ball with a greater release speed, and because the range of a projectile is strongly dependent on the release speed, this bias toward low release angles reduced the optimum release angle to about 30°. Calculations showed that the distance of a throw may be increased by a few metres by launching the ball with a fast backspin, but the ball must be launched at a slightly lower release angle.

## INTRODUCTION

The long throw-in is an important "set play" in soccer, particularly when used as an attacking manoeuvre near to the goal mouth. The farther a player can throw the ball the larger the area in which his/her team mates may receive the ball and the greater the scoring opportunities. To produce a long throw the player must project the ball at high speed and at an appropriate angle with respect to the horizontal. In laboratory studies of the throw-in, male players have recorded release speeds of 12–19 m/s and release angles of 22–40º (Messier and Brody, 1986; Kollath and Schwirtz, 1988; Bray and Kerwin, 2004; Kerwin and Bray, 2004). Release speed is the primary determinant of the range attained by a projectile, and to achieve the greatest possible distance a soccer player should release the ball with the greatest possible speed. Identifying the optimum release angle for a long throw-in is less straightforward. The optimum release angle of a sports projectile is influenced by (1) the physical properties of the projectile, (2) the release conditions, and (3) the anatomical and musculoskeletal constraints of the player's body (de Mestre, 1990; Hubbard, 2000).

### Physical properties of the ball

The competition rules for soccer specify the allowable ranges of the most important physical properties of the ball, such as the weight, circumference, and sphericity (FIFA, 2004). Although the effects of the physical properties of the ball on the throw distance and the optimum release angle have not been investigated, the range of values allowed by the rules is relatively narrow. We therefore expect to see only small changes in the throw distance and optimum release angle arising from changes in the physical properties of the ball. In any case, the physical properties are to a large extent fixed once the ball to be used in the match is selected, and they cannot easily be manipulated by the player in an effort to increase the distance of a throw.

### Release conditions

In a soccer throw-in, the release conditions that affect the optimum release angle are the release speed, release height, and rate of spin. The effects of release speed and release height on the distance and optimum release angle for a moderately aerodynamic projectile such as a soccer ball are well established (de Mestre, 1990). For a ball that is released from a typical release height of 2.3 m, the throw distance increases rapidly with increasing release speed and the throw distance is a maximum at a release angle of about 40º. Changes in release height do not greatly affect the throw distance or the optimum release angle.

The spin imparted to the ball during the release can have a strong influence on the throw distance and the optimum release angle. In a high-speed video study of throws by two male players, Bray and Kerwin (2004) observed backspin at spin rates of between 0.4 and 1.3 rev/s. They presented an aerodynamic model that accurately predicts the flight trajectory of the ball, and this model was used to calculate the effect of backspin on the optimum release angle. For representative values of release speed (18 m/s) and release height (2.3 m), the calculated optimum release angle steadily decreased with increasing backspin, from about 40º for a ball released with zero spin, to about 35º for a backspin of 2 rev/s.

### Anatomical and musculoskeletal constraints

Bray and Kerwin's calculated optimum release angles were substantially greater than the observed release angles for the two players in their study ($26 \pm 3º$ and $32 \pm 4º$). We suspect that closer agreement would have been achieved by accounting for the anatomical and musculoskeletal constraints of the player's body in the calculation of the optimum release angle. Studies of other sports projectile events have shown that an athlete's release speed and release height vary with changes in release angle, and that the relations between these release parameters determine the athlete's optimum release angle (Red and Zogaib, 1977; Linthorne, 2001; Wakai and Linthorne, 2005; Linthorne et al., 2005). For example, in the shot put the musculoskeletal structure of the human body is such that the athlete can produce more throwing force in the horizontal direction than in the vertical direction, and so the maximum release speed that an athlete can attain decreases with increasing release angle (Linthorne, 2001). Because the range of a projectile is strongly dependent on the release speed, even a small dependence of release speed on release angle is sufficient to change the optimum release angle by several degrees. In the shot put, the decrease in speed with increasing release angle reduces the optimum release angle to about 28–36º. Also, a shot putter's release height increases with increasing release angle because of changes in the athlete's body position at the instant of release. However, this relation has only a minor influence on the optimum release angle.

*This study*

The purpose of the present study was to determine the optimum release angle in the long soccer throw-in. A 2-D video analysis of a player was used to obtain mathematical expressions for the relations between release speed and release angle, and between release height and release angle. These expressions were combined with the equations for the flight of a spherical projectile, and the optimum release angle that maximises the distance of the throw was calculated. Simple models of the soccer throw-in were developed to explain the observed relations between release speed, release height, and release angle.

We found that the player produced greater release speeds at lower release angles, and that this relation reduced the optimum release angle to about 30º. Although the player's release height increased with increasing release angle because of changes in his body position at the instant of release, this relation had only a small influence on the optimum release angle. The calculated optimum release angle was in good agreement with the player's preferred release angle. We therefore concluded that our method of calculating the optimum release angle, in which we account for the anatomical and musculoskeletal constraints of the player's body as well as the aerodynamic flight of the ball, produces an accurate prediction of a player's optimum release angle.

In our study the player launched the ball with as little spin as possible so as to reduce the confounding effects of spin on the determination of the optimum release angle. Our calculated optimum projection angle of about 30º therefore only applies to throws in which the player launches the ball with no spin. However, the mechanics of a spinning projectile is well established, and our throw simulations showed that for realistic values of backspin the distance of the throw can be increased by up to a few metres with a slight reduction in the optimum release angle.

## METHODS

In a long throw-in, the throw distance (or horizontal range) $R$ is the horizontal distance the ball's centre of mass travels from the instant of release to the instant of landing (Figure 1). The release parameters that determine the throw distance are the release speed $v$, the release angle $\theta$, the relative release height $h$ (the height difference between the release and the landing), and the spin rate of the ball ($\omega$). To calculate a player's optimum release angle we require the mathematical expressions for the relation between release speed and release angle, $v(\theta)$, and between relative release height and relative release angle, $h(\theta)$. Intervention is therefore required to obtain measurements of a player's release speed and relative release height over a wide range of release angles, rather than just at the player's preferred release angle. The expressions for $v(\theta)$ and $h(\theta)$ may be obtained by fitting mathematical relations to plots of release speed and relative release height as a function of release angle.

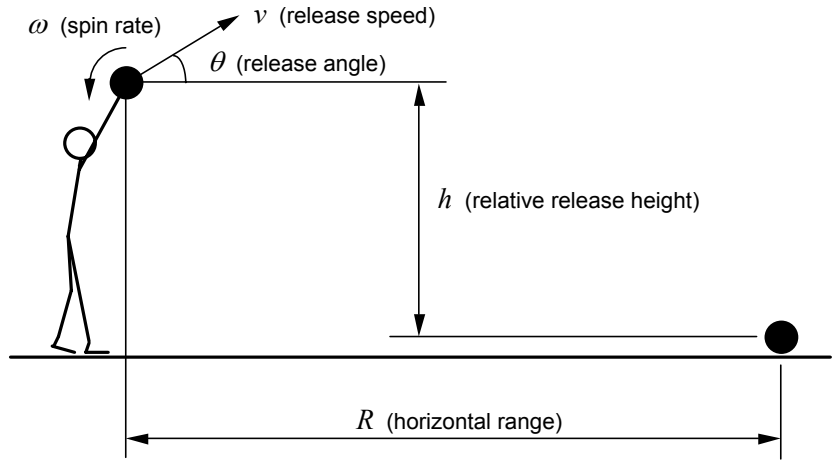

**Figure 1** Diagram of a long soccer throw-in showing the release conditions that determine the horizontal range of the ball.

### Participant and throwing protocol

One collegiate male soccer player (age 21 years; height 1.77 m) was recruited to participate in the study. The study was approved by the Human Ethics Committee of Brunel University, the participant was informed of the protocol and procedures prior to his involvement, and written consent to participate was obtained. The throws were conducted in still air conditions in an outdoor stadium using a FIFA approved match ball (Mitre Ultimatch, size 5). All throws were performed from a flat synthetic surface and the landing area was level with the release surface. The participant wore athletic training clothes and sports shoes. According to FIFA competition rules the player must be facing the field of play when releasing the ball and have part of each foot either on the touchline or on the ground outside the touchline. The player must deliver the ball from behind and over the head with both hands. In this study the participant was instructed to use a self-selected throwing technique that complied with FIFA regulations.

The participant was instructed to perform five maximum-effort throws at his preferred release angle, and then twenty maximum-effort throws at other release angles that ranged from "much higher" to "much lower" than his preferred release angle. The order of the other release angles was altered to preclude any effect resulting from the order. For all throws the participant was instructed to release the ball with as little spin as possible (i.e. $\omega \approx 0$ rev/s). Depending on the intended angle of release, he elected to use a short run-up of 0–4 strides and a delivery stance with the feet from 0 to 0.8 m apart. A unlimited rest interval was given between throws to minimise the effects of fatigue on throwing performance. The throw distance was measured to the nearest 10 cm using a fibreglass tape measure.

*Video analysis*

A Panasonic AG-455 S-VHS video camera operating at 50 Hz was used to record the movement of the ball and player at release. The video camera was mounted on a rigid tripod at a height of 1.8 m, and placed at a right angle to the throw direction about 25 m away from the throw line. The field of view was zoomed to allow the ball to be in the field of view throughout the run-up and delivery and for at least 10 frames after release. The movement space of the video camera was calibrated with three vertical calibration poles that were placed along the midline of the throwing movement and about 3 m apart.

An Ariel Performance Analysis System was used to manually digitise the motion of the ball in the video images. The two-dimensional coordinates of the centre of mass of the ball were calculated from the digitised data using the direct linear transform (DLT) algorithm. Coordinate data were smoothed using a second-order Butterworth digital filter with a cut-off frequency of 10 Hz, and the velocity of the ball's centre of mass was calculated by direct differentiation of the filtered coordinate data. The choice of cut-off frequency was based on a residual analysis (Winter, 1990) and a visual inspection of the power spectra of the coordinate and velocity data. The instant of release was defined as the first frame in which the ball was observed to break contact with the player's hands. The release speed and release angle of the ball were calculated from the horizontal and vertical speed at the instant of release, and the release height was the vertical distance of the centre of mass of the ball relative to the ground.

All digitising was performed by the same operator to maximise the consistency of the dependent variables. In this study the greatest source of uncertainty in the measured values arose from the sampling frequency of the video camera, and this uncertainty was taken as one half the difference between the value at the instant of release and the value at one frame before the instant of release. The calculated uncertainties due to the video sampling rate were about 0.2 m/s for release speed, 0.07 m for release height, and 1.3° for release angle.

*Model of the flight of a soccer ball*

During its flight through the air a projectile may experience aerodynamic effects that arise from the interaction of the projectile with the surrounding air. In the shot put these aerodynamic effects are negligible and range of the shot may be calculated using the equations for a projectile in free flight. In the soccer throw-in, aerodynamic drag and lift can substantially affect the trajectory of the ball, and so a more complex aerodynamic model must be used to calculate the throw distance.

The magnitude of the drag force on a soccer ball depends on the speed of the ball through the air, the cross-sectional area of the ball, and the shape and surface characteristics of the ball. The shape and surface characteristics are accounted for by a scale factor called the drag coefficient, $C_D$. At speeds typical of the soccer throw-in, a soccer ball has a drag coefficient of about $C_D = 0.2$ (Bray and Kerwin, 2004). The trajectory of a soccer ball is also influenced by the rate and direction of its spin. A spinning ball creates a lift force through the Magnus effect. When backspin is applied to the ball the lift force tends to be upwards and the horizontal range of the ball is increased, and when topspin is applied to the ball the lift force tends to be downwards and the horizontal range of the ball is decreased. The magnitude of the lift force depends on the rate of spin, which is reflected in the value of the lift coefficient, $C_L$. For a ball that is projected with zero spin (as in the present study) the lift coefficient is about $C_L = 0$.

We analysed the trajectory of the soccer ball in a rectangular coordinate system where the positive *x*-axis is in the forward horizontal direction and the positive *y*-axis is vertically upwards. The flight trajectory equations of the soccer ball are then (de Mestre, 1990; Bray and Kerwin, 2004)

$$\frac{d^2x}{dt^2} = -kv\left(C_D \frac{dx}{dt} + C_L \frac{dy}{dt}\right) \quad (1)$$

and

$$\frac{d^2y}{dt^2} = kv\left(C_L \frac{dx}{dt} - C_D \frac{dy}{dt}\right) - g, \quad (2)$$

where *v* is the speed of the ball relative to the air, and *g* is the acceleration due to gravity. The constant *k* is given by $k = \rho S/(2m)$, where $\rho$ is the air density (1.225 kg/m³ at sea level and 15°C), *S* is the cross-sectional area of the ball (0.038 m²), and *m* is the mass of the ball (0.43 kg).

If the initial conditions of the ball (i.e. release speed, release height, and release angle) are known, the trajectory of the ball may be computed and the distance of the throw determined. Because the release speed, release height, and release angle are inter-related, we used the measured expressions for $v(\theta)$ and $h(\theta)$ to generate the initial conditions for the flight trajectory equations. The flight trajectory equations are non-linear and so must be computed using numerical methods. In this study we used a technical computing software package (Mathematica; Wolfram Research, Champaign, IL) to calculate the flight trajectories. The calculated throw distance was plotted against release angle, and the optimum release angle was the point on the curve at which the throw distance was greatest.

## RESULTS

The mean values of the throw distance, release speed, release height, and release angle for the throws at the participant's preferred release angle were $17.0 \pm 0.8$ m, $13.4 \pm 0.3$ m/s, $2.23 \pm 0.02$ m and $32.1 \pm 1.6°$ (mean $\pm$ $s$) respectively. These values are similar to those reported in other studies of the soccer throw-in (Messier and Brody, 1986; Kollath and Schwirtz, 1988; Bray and Kerwin, 2004; Kerwin and Bray, 2004). Figures 2 and 3 show the release height and release speed as a function of the release angle.

### *Release height*

A simple anthropometric model of the throw was developed and used to explain the observed increase in release height with increasing release angle. At the instant of ball release the player was standing almost upright with his arms straight and at an angle $\alpha$ to the horizontal (Figure 4). The release height is then given by

$$h_{\text{release}} = h_{\text{shoulder}} + l_{\text{arm}} \sin \alpha, \qquad (3)$$

where $h_{\text{shoulder}}$ is the height of the player's shoulders when standing upright, and $l_{\text{arm}}$ is the length of the player's outstretched arms. The angle of the arms to the horizontal was not the same as the release angle, but there was a strong linear relation between the two angles. The release height may therefore be expressed by

$$h_{\text{release}} = h_{\text{shoulder}} + l_{\text{arm}} \sin (A\theta + \alpha_{\text{o}}), \qquad (4)$$

where $A$ is the rate of increase in the arm angle with increasing release angle, and $\alpha_{\text{o}}$ is the arm angle for a horizontal release. In this study we required an expression for the relative release height, rather than the release height. The relative release height is given by

$$h = h_{\text{release}} - h_{\text{landing}}, \qquad (5)$$

where $h_{\text{release}}$ is the height of the ball at release and $h_{\text{landing}}$ is the height of the ball at landing. When throwing on level ground the landing height is equal to the radius of the ball, and so we obtain

$$h(\theta) = h_{\text{shoulder}} + l_{\text{arm}} \sin(A\theta + \alpha_{\text{o}}) - r_{\text{ball}}, \qquad (6)$$

where $r_{ball}$ is the radius of the ball (0.11 m).

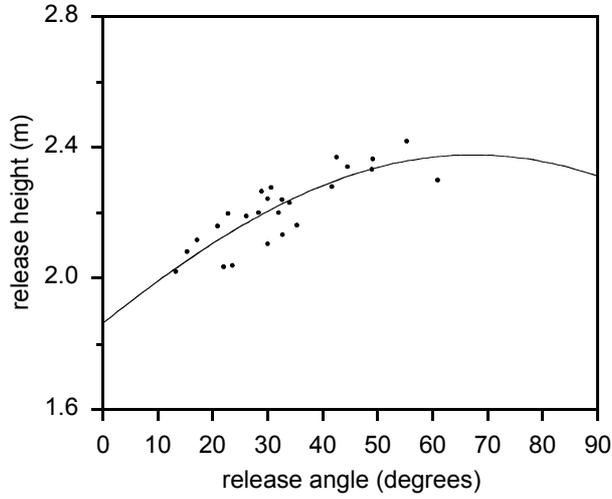

**Figure 2** Release height as a function of release angle for a male player. The fitted curve is from equation (4).

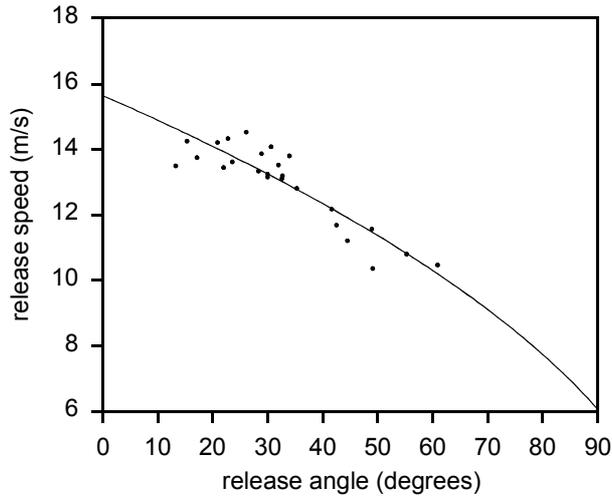

**Figure 3** Release speed as a function of release angle. The fitted curve is from equation (9).

A curve of the form of equation (4) was fitted to the plot of release height as a function of release angle (Figure 2) by selecting values of $h_{shoulder}$, $l_{arm}$, $A$ and $\alpha_o$ using the Levenberg-Marquardt algorithm (Press *et al.*, 1988). However, with four fitted parameters the uncertainties in some of the calculated values were about the same magnitude as the value of the parameter. Better results were obtained with just three fitted parameters ($h_{shoulder}$, $A$ and $\alpha_o$). Video measurements showed that the distance between the player's shoulders and the ball at the instant of release remained approximately constant across all release angles, with an average value of $l_{arm} = 0.80$ m. A curve was again fitted to the plot of release height as a function of release angle. The calculated values and standard errors of the fitted parameters were $h_{shoulder} = 1.58 \pm 0.07$ m, $A = 1.0 \pm 0.3$ and $\alpha_o = 21 \pm 5°$, and the fitted curve for the participant is shown in Figure 2. The calculated shoulder height is in approximate agreement with the player's actual body dimensions (1.45 m).

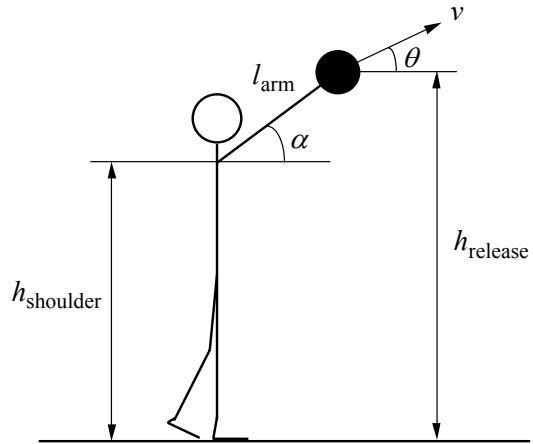

**Figure 4** Anthropometric model of the player at the instant of release. This model was used to explain the observed relation between the release height and release angle (Figure 2).

*Release speed*

The release action in the soccer throw-in is a moderately complex movement. A simple model of the release was devised in an attempt to explain the observed decrease in release speed with increasing release angle (Figure 3). As the player moves the ball back behind his head in preparation for the thrust with the arms, there is a moment when the speed of the ball is close to zero. The throw-in action was reduced to just the thrust phase, from the time the ball is stationary to the instant of release. During the thrust with the arms the player is assumed to apply a constant force $F$ to the ball. This force accelerates the ball along a straight-line path to produce a release angle $\theta$. We assume that the weight of the ball is negligible in comparison to the force exerted by the player on the ball. The release speed is then given by (Linthorne, 2001)

$$v = \sqrt{\frac{2Fl}{m}}, \qquad (7)$$

where $l$ is the length of the acceleration path, and $m$ is the mass of the ball.

Unfortunately this model is too simplistic. The model implies that the release speed is independent of release angle, but our measurements show that release speed decreases with increasing release angle (Figure 3). We therefore modified the model by assuming that the musculoskeletal structure of the human body is such that a player can exert more force on the ball when throwing horizontally than when throwing vertically (Linthorne, 2001). We also assume that the force exerted by the player on the ball decreases linearly with release angle. That is, the force is given by

$$F = F_o - a\theta, \qquad (8)$$

where $F_o$ is the average force exerted on the ball for a horizontal release angle, and $a$ is a constant that characterizes the force decrease with increasing release angle. The constant $a$ is expected to be specific to the player. It should depend on the player's body dimensions, muscle strengths, and throwing technique. The relation between release speed and release angle is then given by

$$v(\theta) = \sqrt{\frac{2(F_o - a\theta)l}{m}}. \qquad (9)$$

A curve of the form of equation (9) was fitted to the plot of release speed as a function of release angle (Figure 3) by selecting values of $F_o$, $l$ and $a$ using the Levenberg-Marquardt algorithm (Press *et al.*, 1988). However, with three fitted parameters the uncertainties in the calculated values were several hundred times the value of the parameter. Better results were obtained with just two fitted parameters ($F_o$ and $a$). Our video measurements showed that the acceleration path length remained approximately constant across all release angle, with an average value of $l = 1.14$ m. A curve was again fitted to the plot of release speed as a function of release angle. The calculated values and standard errors of the fitted parameters were $F_o = 46 \pm 2$ N and $a = 0.44 \pm 0.04$ N/degree, and the fitted curve for the participant is shown in Figure 3. The average force exerted by the player on the ball is much less than the force the player can exert in an isometric contraction. The difference between static and dynamic force production is a well-known phenomenon owing to the force-velocity relation of contracting muscle.

Although equation (9) gives a good fit to the experimental data and the fitted value of $F_o$ appears to have some correspondence to physical reality, the model behind the equation is simplistic. The release force generated by the player is probably not constant, as assumed in the model. The force generated by the player is expected to change throughout the release because of the changing lengths, moment arms, and contraction speeds of the player's muscles (van den Tillaar and Ettema, 2004). Also, we have no experimental evidence that the average force exerted by the player is the same for all release angles. Therefore, the practical value of the simple model used here should not be extended beyond the ability to provide a good fit to the release speed versus release angle data (Figure 3).

*Optimum release angle*

The optimum release angle for the participant was calculated and compared with his preferred release angle. To calculate the optimum release angle the values of $h_{\text{shoulder}}$, $A$ and $\alpha_0$ were substituted into equation (6) and the values of $F_o$ and $a$ were substituted into equation (9). The resulting expressions for $h(\theta)$ and $v(\theta)$ were then used to generate the initial conditions for the flight trajectory equations (equations 1 and 2) for a series of release angles between 0° and 90° in steps of 0.01°. For each release angle the flight trajectory was calculated and the throw distance and flight time were recorded. The calculated throw distance was plotted against release angle, and the optimum release angle was the point on the curve at which the throw distance was greatest.

The flight trajectory calculation required an estimate of the drag coefficient of the ball. In this study we used the measured throw distances to determine the drag coefficient. For each of the 25 throws by the participant, the range of the throw was calculated by using the measured release speed, release height, and release angle as the initial conditions in the flight trajectory model. The drag coefficient in the flight trajectory model was adjusted from 0.10 to 0.30 in increments of 0.01, and the calculated throw distances for each of the 25 throws by the participant were recorded. Best agreement between the calculated throw distances and the measured throw distances was achieved with a drag coefficient of $C_D = 0.25$, and this value was taken as the drag coefficient of the ball in the calculations of the optimum release angle.

For the player in this study, the calculated optimum release angle (29.7° ± 1.5°) was in good agreement with the participant's preferred release angle (32.1 ± 0.8°). The calculated optimum release angle was insensitive to the value of the drag coefficient that was used in the flight trajectory model. A drag coefficient of $C_D = 0.15$ gave an optimum release angle of 30.1° and drag coefficient $C_D = 0.35$ gave an optimum release angle of 29.4°. Therefore, we conclude that our method, in which the measured relations for $v(\theta)$ and $h(\theta)$ are combined with the equations that describe the flight trajectory of ball, produces an accurate calculation of a player's optimum release angle.

The combined effects of $v(\theta)$ and $h(\theta)$ on the optimum release angle are illustrated in Figure 5. For a non-aerodynamic projectile that is released from ground level, the optimum release angle is 45°. However, this optimum release angle is only appropriate if the athlete can produce the same release speed at all release angles. The musculoskeletal structure of the human body is such that a soccer player is able to produce a greater release speed when releasing the ball a low release angle. Because the range of a projectile is strongly dependent on the release speed, the bias towards low release angles reduces the optimum release angle by about 10°. Also, in a soccer throw-in the ball is released from about 2 m above the landing, and this release height increases slightly with increasing release angle because of changes in the player's body position at the instant of release. However, the player's height-angle relation has a relatively small effect on the optimum release angle; it reduces the optimum release angle by only about 4°. A soccer ball is a moderately aerodynamic projectile and it experiences substantial aerodynamic drag during its flight through the air. Although aerodynamic drag reduces the throw distance by a few metres, it reduces the optimum release angle by only about 2°.

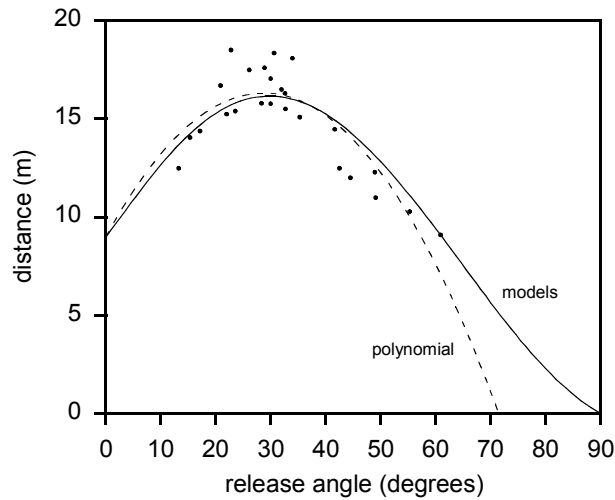

**Figure 5** Measured throw distance as a function of release angle. Also shown is the calculated throw distance (solid line) and a polynomial fit to the throw distance data (dashed line). The optimum release angle for this player is about 30°.

An alternative method of identifying the optimum release angle for a player is to fit a regression curve directly to the measured throw distance versus release angle data (Figure 5). A second-degree polynomial was found to be the most appropriate fit as it produced the lowest coefficient of variation in the regression equation coefficients. The calculated optimum release angle obtained using this method was very close to that obtained using equations (6) and (9). However, this direct method of determining the optimum release angle has a greater uncertainty in the calculated value (28.7 ± 8.5°) and does not shed light on the factors that determine the player's optimum release angle.

An important practical result from our study is that projecting the ball at the optimum release angle is not very important in a producing a long throw. From Figure 5 we see that the throw distance is not sensitive to release angle, and so relatively large errors in release angle can be tolerated. For the player in this study the release angle needed to be within about 7° of the optimum release angle for the throw to be within 0.5 m (3%) of the maximum achievable distance.

## DISCUSSION and IMPLICATIONS

### Limitations of the study

The use of a 50 Hz video camera to measure the release variables was not a significant limitation of this study. The variability in the participant's throws was greater than the measurement uncertainties introduced through the sampling rate of the video camera. Therefore, using a video camera with a sampling rate of 100 or 200 Hz would not have substantially reduced the uncertainty in the mathematical expressions for $v(\theta)$ and $h(\theta)$, and would not have produced a more accurate estimate of the optimum release angle.

Although several simplifying assumptions were introduced to the models that were used to determine $v(\theta)$ and $h(\theta)$, the conclusions of the study are not affected by these simplifications. The mathematical expressions used for $v(\theta)$ and $h(\theta)$ produced good fits to the experimental data (Figures 2 and 3), and gave an accurate calculation of the optimum release angle. As an alternative to the models, we obtained expressions for $v(\theta)$ and $h(\theta)$ by fitting polynomial equations to the release speed and release height data. For both $v(\theta)$ and $h(\theta)$, a first-degree polynomial produced the lowest coefficient of variation in the regression equation coefficients. The calculated optimum release angle obtained using the polynomials was similar (29.3 ± 1.5°) to that obtained using equations (6) and (9).

### Flight time

Launching the ball at the release angle that maximises the throw distance may not be the most appropriate strategy. In the attacking long throw-in the player would like to maximise the element of surprise by passing the ball to his/her team mate in as short a time as possible. At release angles near to the optimum release angle (30º) the flight time decreases with decreasing release angle (Figure 6). Therefore, if the player launches the ball a few degrees below the optimum release angle, the throw distance will be essentially the same but the flight time of the throw will be reduced. For the player in this study the flight time decreases at a rate of about 0.025 s per 1° decrease in release angle. If he throws the ball at 4° below the optimum release angle, the throw distance will be almost the same as the maximum possible distance (99%) but the flight time of the throw will be reduced by about 0.1 s (7%). If he decides to sacrifice a little more distance (6%) by launching the ball at 10° below the optimum release angle, he can gain a time advantage of about 0.3 s (20%). A reduction in flight time of 0.1–0.3 s might be the decisive factor in turning a long throw-in into a goal scoring opportunity.

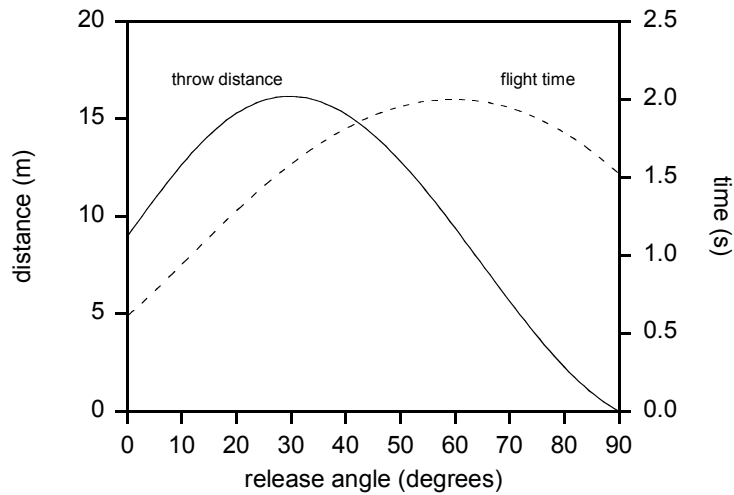

**Figure 6** Throw distance and flight time as a function of release angle.

*Inter-individual differences*

In this study we investigated the optimum release angle in only one male player. In a study of the shot put, Linthorne (2001) found that the optimum release angle was unique to the athlete. The five athletes in his study had optimum release angles that differed by 6° because of inter-individual differences in the shape of the speed-angle curve. We suspect that in the soccer throw-in the optimum release angle may also be unique to the player. Inter-individual differences in body size, muscular strength and elasticity, and throwing technique may produce differences in the shape of the speed-angle curve, and hence result in different optimum release angles.

    The two players in the study by Bray and Kerwin (2004) were more proficient in the long throw-in than the player in our study. For the two players in Bray and Kerwin's study, we estimated the shape of the speed-angle relation that would produce throw distances and optimum release angles equal to their observed values (28.9 m and 26º for Player A; 21.1 m and 32º for Player B). In our model of the throw-in (equation 9), the throw distance is mostly determined by the average force exerted by the player on the ball ($F_o$), and the optimum release angle is mostly determined by the rate of force decrease with increasing release angle ($a$). The greater the average force the longer the throw distance, and the greater the rate of force decrease the lower the optimum release angle. In

comparison to the player in the present study, we calculate that Player A had a much greater throwing force ($F_o$ = 125 N) and a greater rate of force decrease ($a$ = 1.35 N/degree), whereas Player B had a slightly greater throwing force ($F_o$ = 62 N) and a similar rate of force decrease ($a$ = 0.44 N/degree).

In the long soccer throw-in it is much more important for a player to attain a high release speed than to throw at the optimum release angle. The range of a moderately aerodynamic projectile is approximately proportional to the square of the release speed. The implication is that to achieve longer throws the player should work on developing muscular strength and on improving his/her throwing technique. Improvements in strength and/or technique are expected to shift the speed-angle relation (Figure 3) upwards, hence producing a greater maximum throw distance at a similar optimum release angle.

### *Ball size and mass*

We used the flight trajectory equations (equations 1 and 2) and the measured $v(\theta)$ and $h(\theta)$ relations to investigate the effects of the physical properties of the ball on the maximum throw distance and the optimum release angle. FIFA (2004) competition rules state that the circumference of the ball must be between 68 and 70 cm, and the weight of the ball must be between 410 and 450 g. A larger diameter ball experiences a greater aerodynamic drag during its flight through the air, and so the maximum throw distance is shorter and the optimum release is lower than for a small ball. For a heavy ball the relative effect of aerodynamic drag compared to that of gravity is less, and so the maximum throw distance and optimum release angle are greater than for a light ball. However, the effects of the permitted variations in the circumference and mass of the ball are very small. The differences in the throw distance and optimum release angle between the upper and lower limits of ball circumference are only 0.12 m and 0.05°, and the differences between the upper and lower limits of ball mass are only 0.19 m and 0.08°. Although the preceding calculations are specific to the player in our study, we do not expect substantially different values for other players. The calculated effects are so small that they would be difficult to verify experimentally, even with a ball-launching machine. We conclude that variations in ball mass and circumference have no practical implications in the long throw-in.

### *Passes to head-height and chest-height*

In this study, we assumed that the ball lands on the ground and that the player wishes to maximise the horizontal distance that the ball travels from release to landing. In practice, a player may sometimes wish to pass the ball to a player so that s/he receives the ball at chest-height or head-height. We repeated our calculations of the optimum release angle so that the ball was received by the player at a range of heights from ground level to 2.0 m above the ground (Figure 7). Receiving the ball above ground level reduces the throw distance and increases the optimum release angle. If the player receives the ball at chest height (about 1.3 m), the throw distance is reduced by about 1.3 m (8%) and the optimum release angle is increased by about 2.3°. For a pass to head-height (about 1.8 m) the throw distance is reduced by about 1.9 m (12%) and the optimum release angle is increased by about 3.4°.

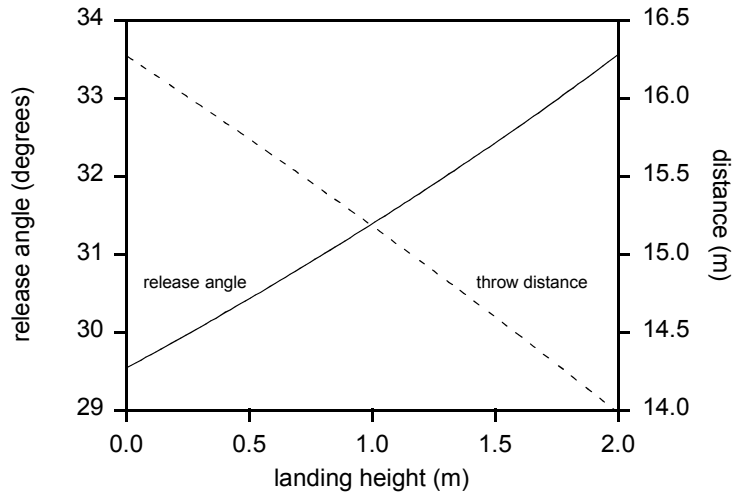

**Figure 7** Calculated effect of the landing height on the throw distance and optimum release angle.

*Backspin*

In our study the player launched the ball with as little spin as possible so as to reduce the confounding effects of spin on the determination of the optimum release angle. The calculated optimum projection angle for the player of about 30º therefore only applies to throws in which the ball is launched with no spin. However, the influence of spin on the trajectory of a spherical projectile such as a soccer ball is well established, and so we performed calculations to determine the effects of backspin on the player's maximum throw distance and the optimum release angle.

A ball that is launched with backspin produces a lift force that tends to increase the distance of the throw. The magnitude of the lift force depends on the rate of spin and is reflected in the value of the lift coefficient, $C_L$. Video analysis of the flight trajectories of a soccer ball indicate that the lift coefficient increases exponentially with increasing spin rate, from $C_L = 0$ for no spin, up to a limiting value of about $C_L = 0.25$. Here, we assume a relation given by $C_L = -0.25\,e^{-0.5\omega} + 0.25$, where $\omega$ is the spin rate in revolutions per second (Carré *et al.*, 2002; Bray and Kerwin, 2004). A spinning ball also has a slightly greater drag coefficient than a non-spinning ball. For a spinning soccer ball the drag coefficient is expected to increase at a rate of about 0.01 per 1 rev/s increase in spin rate (Smits and Smith, 1994).

The calculated effects of backspin on the player's maximum throw distance and optimum release angle are shown in Figure 8. Most of the effects are due to the increase in the lift coefficient. The increase in lift coefficient tends to produce a longer throw and a lower optimum release angle, whereas the increase in drag coefficient tends to produce a shorter throw and a lower optimum release angle. A spin of 3 rev/s is probably close to the maximum that a player can achieve in a throw-in. If a player were to achieve a spin rate of 3 rev/s the throw distance would be increased by about 1.2 m (8%) and the optimum release angle would be lowered by about 3°. In our calculations the throw distance reaches a maximum at a spin rate of about 5 rev/s. At higher spin rates the gain in distance due to the greater lift is outweighed by the loss in distance due to greater drag.

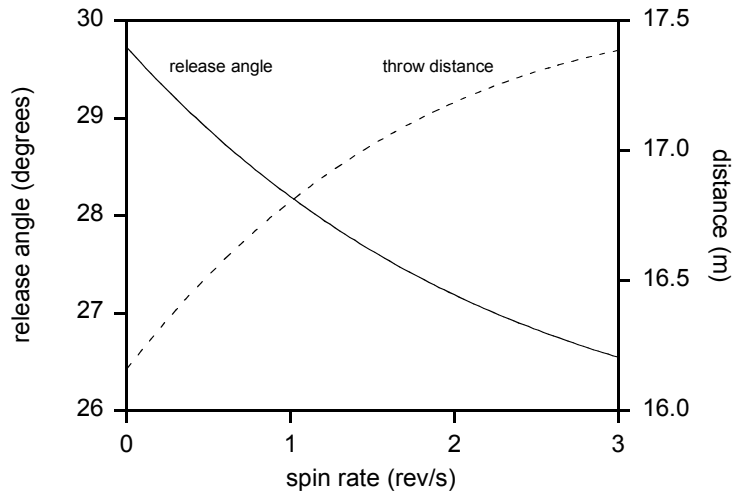

**Figure 8** Calculated effect of backspin on the throw distance and optimum release angle.

## CONCLUSIONS AND RECOMMENDATIONS

This study showed that the optimum release angle in the long soccer throw-in is considerably less than 45°. The release speed a player can generate increases at lower release angles, and this bias reduces the optimum release angle to about 30°. The height difference between release and landing has only a small effect on the optimum release angle. The optimum release angle is probably slightly different for each player, and may depend on the player's body size, muscular strength, and throwing technique. To produce a longer throw we recommend that the player work on increasing the release speed by developing explosive strength in the muscles used in the throwing movement, and by improving his/her throwing technique.

It is not essential to launch the ball at precisely the optimum release angle as deviations of several degrees do not substantially reduce the distance of the throw. The distance achieved in a long throw is greatest when the player receives the ball at ground level. When a throw is received at chest-height or head-height the throw distance is a little less and the ball should be launched a few degrees higher.

In an attacking long throw, launching the ball for maximum distance may not be the most appropriate strategy. We recommend that the player should deliberately launch the ball a few degrees below the optimum release angle. This will reduce the flight time of the throw without substantially reducing the throw distance, and hence provide the attacking team with a greater element of surprise. We also recommend that the player should launch the ball with a high backspin. This will increase the distance of the throw, as long as the player's technique for producing backspin does not reduce the release speed. The player should reduce the release angle by a few degrees if s/he launches the ball with substantial backspin.